\documentclass[aps,prl,twocolumn, superscriptaddress]{revtex4}

\usepackage{graphicx}

\begin{document}


\title{The mode-locking transition of random lasers}



\author{Marco Leonetti}
\affiliation{Instituto de Ciencia de Materiales de Madrid (CSIC) and Unidad Asociada CSIC-UVigo, Cantoblanco 28049 Madrid Espa\~{n}a.}

\author{Claudio Conti}
\affiliation{Dep. Molecular Medicine and CNR-ISC Dep. Physics, University Sapienza, P.le Aldo Moro 5, I-00185, Roma Italy}

\author{Cefe Lopez}
\affiliation{Instituto de Ciencia de Materiales de Madrid (CSIC) and Unidad Asociada CSIC-UVigo, Cantoblanco 28049 Madrid Espa\~{n}a.}


\begin{abstract}
The discovery of the spontaneous mode-locking of lasers, i.e., the synchronous oscillation of electromagnetic modes in a cavity, has been a milestone of photonics allowing the realization of oscillators delivering ultra-short pulses. This process is so far known to occur only in standard ordered lasers with meter size length and only in the presence of a specific device (the saturable absorber).
Here we demonstrate that mode-locking can spontaneously arise also in random lasers composed by micron-sized laser resonances dwelling in intrinsically disordered, self-assembled clusters of nanometer-sized particles.
Moreover by engineering a novel mode-selective pumping mechanism we show that it is possible to continuously drive the system from a configuration in which the various excited electromagnetic modes oscillate in the form of several, weakly interacting, resonances to a collective strongly interacting regime. By realizing the smallest mode-locking device ever fabricated, we open the way to novel generation of miniaturized and all-optically controlled light sources.
\end{abstract}

\date{\today}
\maketitle

Random lasers\cite{Wiersma_Rew} (RLs) are made by disordered highly scattering materials able to amplify light when externally pumped. The simultaneous presence of structural disorder and nonlinearity makes these devices a fertile ground to connect photonics with advanced theoretical paradigms\cite{Kaiser_CA} like chaos\cite{Mujumdar_Chaotic}, non Gaussian statistics\cite{Lepri_Wiersma_Levi}, complexity\cite{Leuzzi09} and also the physics of Bose Einstein condensation\cite{Conti_condensation}. Historically there has been a bridge in the RL interpretation. In pioneering experiments a smooth, single-peaked emission was produced by pumping finely ground laser crystals\cite{Gouedard_RL_by_powder}, or titania particles dispersed in a dye-doped solution\cite{Letokhov_NRA, Lawandy_Nature}. This phenomenon has been dubbed RL with incoherent feedback (IFRL) because it may be explained in the framework  of the diffusion approximation \cite{Wiermsa_Diff_Gain} that neglects interference and treats light rays as the trajectories of random walking particles. However this theoretical framework does not explain another kind of RL exhibitting sub-nanometre sharp spectral peaks\cite{Cao_RL_Action_Semiconductor, Lagendijk_spatial_extent_RL, VanderMolen_Several_RL} associated with high-Q resonances\cite{Conti08PhC, Cao_spatial_confinement, Fallert_coexistence_nature, Tureci:08} and labeled resonant feedback random laser(RFRL).

Standard multimode lasers without disorder and characterized by equispaced resonances may be driven to a synchronous regime through the so called mode-locking transition\cite{Haus_Mode_Loking, Kutz_MLSL}, which so far has only been shown to occur spontaneously in the presence of a saturable absorber and allows to generate  ultra-short light pulses  \cite{Gordon_Fischer_PRL_2002, Picozzi_Condesation}. We show that the same transition occurs in RLs and allows to lock modes of a RFRL casting its emission in the typical IFRL spectrum and demonstrating the inherently coherent nature of the random lasing phenomenon.

\begin{figure}[h!]\label{fig:figure1}
\includegraphics[width=\columnwidth]{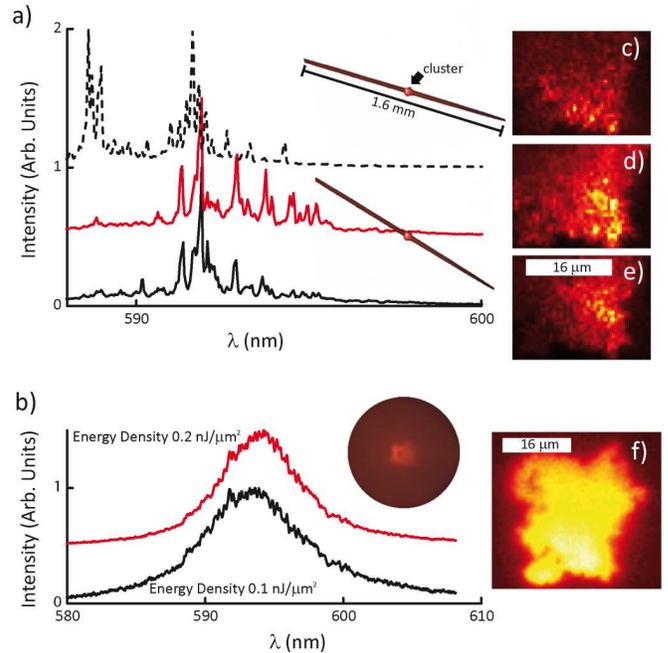}
\caption{The two random lasing regimes.  Panel a) shows three normalized spectra each obtained by averaging 100 single shots by pumping a stripe shaped area ( 1.6 mm long). Top and bottom lines are retrieved for a stripe of the same thickness (16 $\mu m$)  but different orientation (15 $^\circ$ tilt). Middle line has the same orientation of the bottom graph but doubled thickness. Panel b) shows the spectrum for a disk shaped pumping (diameter 1mm) for two different pump densities. The insets show sketches of the pumping areas. Panels c), d), e) and f) show the emitted intensity distribution corresponding to the lines on the left. Pictures are retrieved by optical imaging of the RL emission obtained whit a pumping energy density of 0.1 nJ/$\mu$m$^2$. The length of the white bar in figures e) and f) is 16$\mu$m.}
\end{figure}

The system we consider is an isolated micrometer sized cluster of titania nanoparticles immersed in a rhodamine dye solution (see supplementary information (SI)). In our novel setup we use the amplified spontaneous emission (ASE) from the surrounding dye to pump the cluster. The the ASE areas are defined by shaping the beam of an external (solid state) laser with the help of a reflective spatial light modulator.

\begin{figure}[h!]\label{fig:figure2}
\includegraphics[width=\columnwidth]{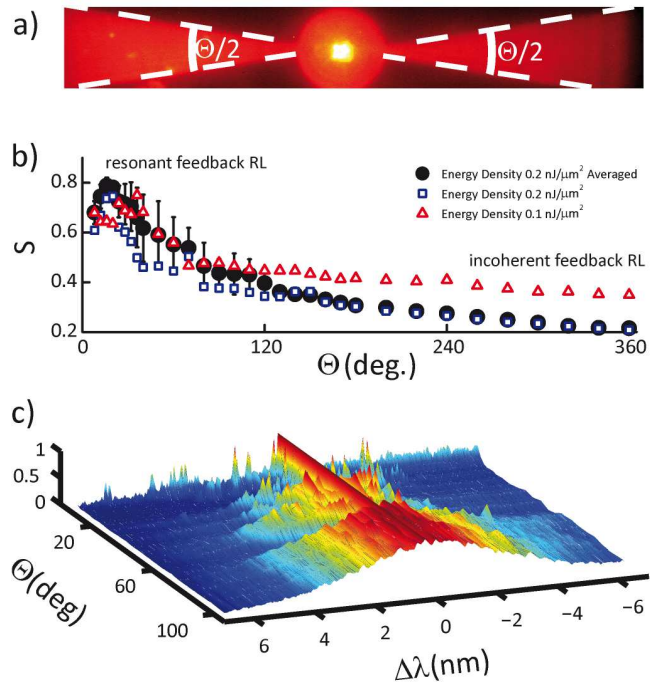}
\caption{Cake pumping. Panel a) depicts the cluster and the surrounding pumped area for $\Theta=36^\circ$. Panel b) shows $S$ as a function of $\Theta$. Squares and triangles correspond to different pump energies for  the cluster C1. Circles corresponds to the average of  5 measures from different clusters. The 3 dimensional graph on panel c) shows normalized spectra ( average over 100 shots with energy density 0.2 nJ/$\mu$m$^2$ ), for different $\Theta$. Spectra are arbitrarily shifted in frequency to superimpose intensity maxima.  $\Delta\lambda$ is the wavelength shift from the most intense peak.}
\end{figure}

We first consider a stripe-shaped ASE pump area of 1.6 mm length (see figure 1a) and $16 \mu$m width to obtain a quasi-one dimensional area that acts as a strongly directional source of ASE hitting the sample which is located in the centre of the stripe. The resulting spectra display sharp peaks. Figure 1a shows the average spectra over 100 pump pulses ("shots") obtained by collecting light emitted off-plane from the center of the cluster. Notably the spectral position of the peaks remain unchanged from shot to shot. Dashed and continuous black lines in figure 1a correspond to stripes differing by a rotation of 15$^\circ$ (see insets). Similar results are obtained for a 32$\mu$m wide stripe (red line in figure 1a). Changing the stripe orientation activates different sets of modes, as revealed by the change in the peaks positions. Panels c,d and e show the spatial intensity distribution corresponding to averaged spectra in figure 1a; panels 1c and 1e correspond to different stripe orientations and display uncorrelated intensity distributions; conversely all the spots in 1e are also present in panel 1d that corresponds to the same stripe orientation but has larger width (red and black continuous lines in figure 1a). Hence the stripe orientation affects spatial distribution of intensity and selects the set of activated modes. Figure 1b shows the measured spectra when the cluster (sample "C1") is placed in the center of a circular pump spot (diameter 1mm) and no directionality is present. In this configuration spectra are smooth and narrow when increasing the energy, while the spatial intensity is homogeneously distributed (panel 1f). Having established that we can selectively excite different modes we proceed to study the effect of the number of the input k-vectors on the RL emission properties.

To study the transition from RFRL to a IFRL we engineered a more complex pumping design (see Figure 2a). An area centered on the cluster is shone consisting of a central disk, 150 $\mu m$ in diameter (to assure homogeneous pumping even to the largest clusters) and two symmetrical wedges of much larger radius ($1 mm$ in diameter) of variable orientation and aperture angle ($\Theta/2$) that serve as ASE launch pad. A single wedge configuration leads to the same results but  proved to be hydrodynamically less stable. The central circle places the cluster barely below lasing threshold preparing the system for lasing once the wedge's ASE is turned on. The angular aperture $\Theta$ controls the number of k-vectors with which ASE is produced and therefore controls the number of modes expected to be excited. Spectra observed for small $\Theta$ ($\sim$10$^\circ$) display several very narrow ( $\sim0.05 nm$) peaks whereas large $\Theta$ ($\sim$100$^\circ$) produces a single and smooth RL lineshape ( $\sim4 nm$) as shown in figure 3 below.

To classify a RL into the IFRL or RFRL categories we measure its \emph{spikiness} by analyzing the Fourier Transform power spectrum (FTS) of the emission spectrum. We define $S$  as the high-frequency fraction of the total FTS area, i.e. the spectral power above a frequency threshold, see SI). Figure 2b shows $S$ versus $\Theta$ at different pump energies for sample C1 (squares and triangles)  and  averaged over 5 different clusters at high energies (circles). All curves display the same trend evidencing a transition: after a rapid growth corresponding to an increase in fluence and number of excited modes (appearing on a smooth fluorescence spectrum), $S$ reaches a maximum (RFRL regime) and then the spectrum becomes smoother as $\Theta$ grows, multiple peaks merge until an IFRL-like emission is achieved (see panel 2c). Note that smoothing at high $\Theta$ is not due to averaging because sharp peaks are absent in single shot spectra too.

$\Theta$ also affects the inter-mode spectral correlation: in figure 3c we show intensities for a random pair of peaks ($I(\lambda_1=597.2 nm)$  and $I(\lambda_2=596.7 nm)$) for $\Theta$=18$^\circ$ for 100 shots revealing an uncorrelated regime (average spectra reported in figure 3a). For $\Theta$=360$^\circ$ the sublte features present on top of the otherwise smooth spectrum (figure 3b) are repeatable from shot to shot (thus characteristic of the considered cluster) and the intensities $I(\lambda_3=598.4 nm)$  and $I(\lambda_4=598.7 nm)$ are strongly correlated (figure 3d). Figure 3e shows the average Pearson correlation $C$ (see SI)  obtained from all possible pairs among the 15 most intense peaks (105 pairs) versus $\Theta$ for sample C1. The onset of a strongly correlated regime is obtained for $\Theta\cong 120^\circ$. The same transition has been observed in all the samples considered, revealing a universal trend in which C reaches unity (within experimental error) when $\Theta>$180$^\circ$. Measurements at fixed pumping volume enable us to exclude the influence of the ASE noise (details in SI).

In former experiments on RFRL a tightly focused pump spot was used to excite a limited number of modes thus obtaining a spectral emission displaying narrow spikes \cite{Cao_spatial_confinement, PhysRevA.81.043830}. In our approach, for small $\Theta$, we select modes that are strongly coupled with a directional input but dwell at distant positions (see figures 1c, 1d and 1e); in the absence of spatial overlap their mutual interaction is negligible, and the spectra obtained feature narrow peaks with limited correlation (figure 3e for low $\Theta$). Conversely, for large $\Theta$, we excite a large number of spatially overlapped resonances, which result in a strongly correlated emission (figure 3e for large $\Theta$), and a spatially uniform intensity distribution without hot spots (figure 1f), due to pronounced interaction between the modes.

\begin{figure}\label{fig:figure3}
\includegraphics[width=\columnwidth]{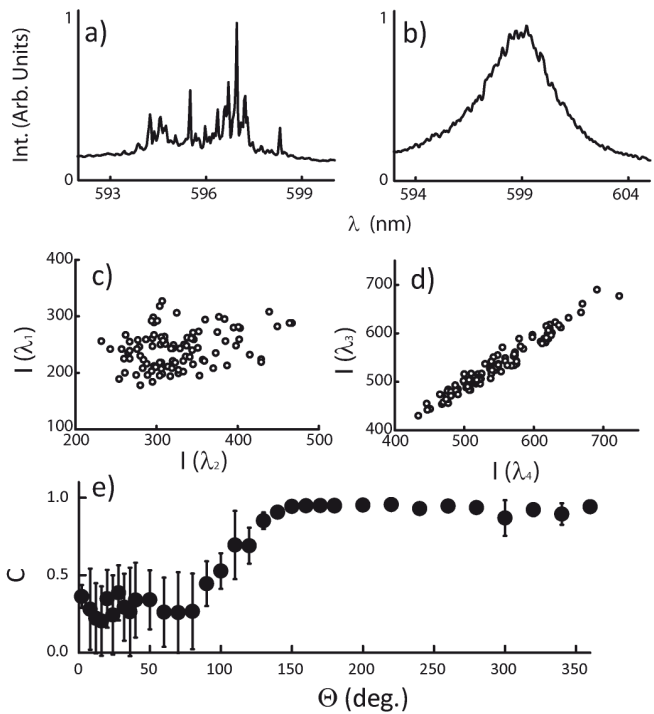}
\caption{The onset of a correlated random laser.  Panels a) and b) report average spectra from cluster C1 for $\Theta$=18$^\circ$ and $\Theta$=360$^\circ$ respectively. In panel c) and d) 100 intensity pairs  (for the chosen wavelengths see text)for 100 single shot spectra are shown for $\Theta$=18$^\circ$ and $\Theta$=360$^\circ$ respectively. Panel e) reports the correlation $C$ averaged over all possible combinations of the 15 most intense peaks versus $\Theta$. Error bars represent statistical errors from all the 105 pairs.}
\end{figure}

We reproduced these results within the framework of coupled mode theory (CMT\cite{Leuzzi09, Conti_condensation, Haus_Mode_Loking}), by considering a set of $N = 50$ modes at different frequencies\cite{PhysRevLett.90.203903}, subject to mode repulsion \cite{PhysRevB.67.161101,PhysRevLett.98.143901} and excited in random initial conditions by an external pump pulse (see SI). In our model the role of $\Theta$ is played by the variable 2$\times$$n_c$ that is the number of resonances to which every mode couples.

Figure 4 reports the result of our CMT calculations: we show average spectra in panels 4a for $n_c$=0 presenting sharp peaks and resembling an RFRL and 4b for $n_c$=10 in which  an IFRL-like emission including small features on top is retrieved. The difference between the two regimes becomes manifest in the time evolution of the modes: while the phases of the 50 weakly coupled ($n_c$=3) modes (figure 4c) oscillate uncorrelated, they begin to synchronize as the coupling increases (panel 4d $n_c$=6) and finally the mode-locked regime is found for $n_c$=10 (figure 4e ). Note that phases are significative only in the time window where the pump pulse is present (range [0.1,0.4] in the figure, details in SI). The numerically retrieved collective parameters $C$ and $S$ (reported in figure 4f respectively as full squares and open triangles as a function of $n_c$), agree with the experimental results.

\begin{figure}\label{fig:figure4}
\includegraphics[width=\columnwidth]{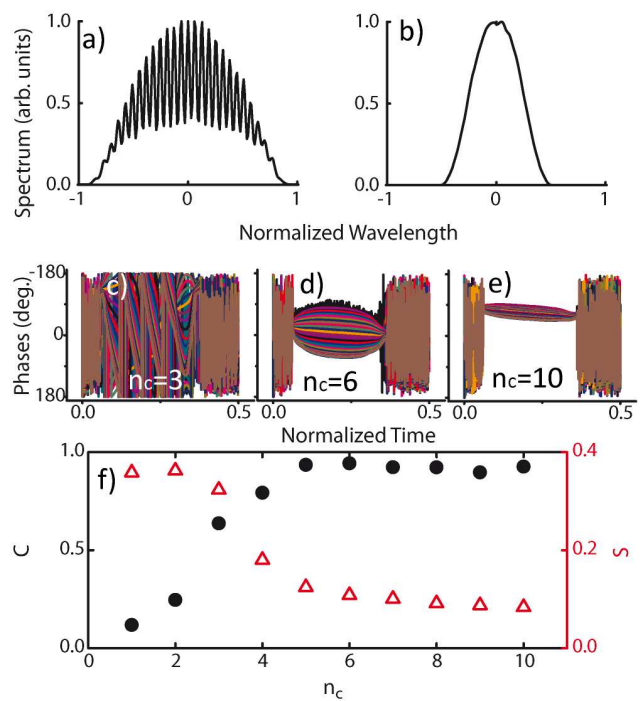}
\caption{Results from numerical CMT calculations. Panel a) and b) display the spectra obtained from 50 modes for $n_c$=0 and $n_c$=10 respectively. Panel c), d) and e) represent phases plotted versus time for $n_c$=3, $n_c$=6 and $n_c$=10. Panel e) reports the numerically calculated $C$ (squares) and $S$ (triangles) as a function of $n_c$.}
\end{figure}

In conclusion, we developed a new mode-locking technique based on a pumping scheme that enables to select the number of activated modes in a random laser. We demonstrate that RLs may be prepared in two distinct regimes, which we can turn on by controlling the shape of the pumping area. When pumping is nearly unidirectional, few (low interacting) modes are turned on and appear as sharp, uncorrelated peaks in the spectrum. By increasing the number of spatial directions hitting the lasing cluster, many resonances intervene generating a smooth emission spectrum with a high degree of correlation, corresponding to the onset of a mode-locked collective state. Only by assuming a transition from a set of nearly independent resonances to a regime with strongly interacting and mode-locked resonances, it is possible to explain the passage from uncorrelated spectral peaks to strongly correlated collective line narrowing. This experiment not only unveils the intimate and unique nature of random lasers and the universality of mode locking process paving the way for a new generation of miniaturized optical devices with engineered and tunable spectral emission, but also lays the foundations for a bridge between disordered photonics and the statistical physics of complex systems.



%


%
%
%


\begin{thebibliography}{25}
\expandafter\ifx\csname natexlab\endcsname\relax\def\natexlab#1{#1}\fi
\expandafter\ifx\csname bibnamefont\endcsname\relax
  \def\bibnamefont#1{#1}\fi
\expandafter\ifx\csname bibfnamefont\endcsname\relax
  \def\bibfnamefont#1{#1}\fi
\expandafter\ifx\csname citenamefont\endcsname\relax
  \def\citenamefont#1{#1}\fi
\expandafter\ifx\csname url\endcsname\relax
  \def\url#1{\texttt{#1}}\fi
\expandafter\ifx\csname urlprefix\endcsname\relax\def\urlprefix{URL }\fi
\providecommand{\bibinfo}[2]{#2}
\providecommand{\eprint}[2][]{\url{#2}}

\bibitem[{\citenamefont{Wiersma}(2008)}]{Wiersma_Rew}
\bibinfo{author}{\bibfnamefont{D.~S.} \bibnamefont{Wiersma}},
  \bibinfo{journal}{Nat Phys} \textbf{\bibinfo{volume}{4}},
  \bibinfo{pages}{359} (\bibinfo{year}{2008}).

\bibitem[{\citenamefont{Froufe-P\'erez
  et~al.}(2009)\citenamefont{Froufe-P\'erez, Guerin, Carminati, and
  Kaiser}}]{Kaiser_CA}
\bibinfo{author}{\bibfnamefont{L.~S.} \bibnamefont{Froufe-P\'erez}},
  \bibinfo{author}{\bibfnamefont{W.}~\bibnamefont{Guerin}},
  \bibinfo{author}{\bibfnamefont{R.}~\bibnamefont{Carminati}},
  \bibnamefont{and} \bibinfo{author}{\bibfnamefont{R.}~\bibnamefont{Kaiser}},
  \bibinfo{journal}{Phys. Rev. Lett.} \textbf{\bibinfo{volume}{102}},
  \bibinfo{pages}{173903} (\bibinfo{year}{2009}).

\bibitem[{\citenamefont{Mujumdar et~al.}(2007)\citenamefont{Mujumdar,
  T{\"u}rck, Torre, and Wiersma}}]{Mujumdar_Chaotic}
\bibinfo{author}{\bibfnamefont{S.}~\bibnamefont{Mujumdar}},
  \bibinfo{author}{\bibfnamefont{V.}~\bibnamefont{T{\"u}rck}},
  \bibinfo{author}{\bibfnamefont{R.}~\bibnamefont{Torre}}, \bibnamefont{and}
  \bibinfo{author}{\bibfnamefont{D.~S.} \bibnamefont{Wiersma}},
  \bibinfo{journal}{Phys. Rev. A} \textbf{\bibinfo{volume}{76}},
  \bibinfo{pages}{033807} (\bibinfo{year}{2007}).

\bibitem[{\citenamefont{Lepri et~al.}(2007)\citenamefont{Lepri, Cavalieri,
  Oppo, and Wiersma}}]{Lepri_Wiersma_Levi}
\bibinfo{author}{\bibfnamefont{S.}~\bibnamefont{Lepri}},
  \bibinfo{author}{\bibfnamefont{S.}~\bibnamefont{Cavalieri}},
  \bibinfo{author}{\bibfnamefont{G.-L.} \bibnamefont{Oppo}}, \bibnamefont{and}
  \bibinfo{author}{\bibfnamefont{D.~S.} \bibnamefont{Wiersma}},
  \bibinfo{journal}{Phys. Rev. A} \textbf{\bibinfo{volume}{75}},
  \bibinfo{pages}{063820} (\bibinfo{year}{2007}).

\bibitem[{\citenamefont{Leuzzi et~al.}(2009)\citenamefont{Leuzzi, Conti, Folli,
  Angelani, and Ruocco}}]{Leuzzi09}
\bibinfo{author}{\bibfnamefont{L.}~\bibnamefont{Leuzzi}},
  \bibinfo{author}{\bibfnamefont{C.}~\bibnamefont{Conti}},
  \bibinfo{author}{\bibfnamefont{V.}~\bibnamefont{Folli}},
  \bibinfo{author}{\bibfnamefont{L.}~\bibnamefont{Angelani}}, \bibnamefont{and}
  \bibinfo{author}{\bibfnamefont{G.}~\bibnamefont{Ruocco}},
  \bibinfo{journal}{Phys. Rev. Lett.} \textbf{\bibinfo{volume}{102}},
  \bibinfo{eid}{083901} (\bibinfo{year}{2009}).

\bibitem[{\citenamefont{Conti et~al.}(2008)\citenamefont{Conti, Leonetti,
  Fratalocchi, Angelani, and Ruocco}}]{Conti_condensation}
\bibinfo{author}{\bibfnamefont{C.}~\bibnamefont{Conti}},
  \bibinfo{author}{\bibfnamefont{M.}~\bibnamefont{Leonetti}},
  \bibinfo{author}{\bibfnamefont{A.}~\bibnamefont{Fratalocchi}},
  \bibinfo{author}{\bibfnamefont{L.}~\bibnamefont{Angelani}}, \bibnamefont{and}
  \bibinfo{author}{\bibfnamefont{G.}~\bibnamefont{Ruocco}},
  \bibinfo{journal}{Phys. Rev. Lett.} \textbf{\bibinfo{volume}{101}},
  \bibinfo{pages}{143901} (\bibinfo{year}{2008}).

\bibitem[{\citenamefont{Gouedard et~al.}(1993)\citenamefont{Gouedard, Husson,
  Sauteret, Auzel, and Migus}}]{Gouedard_RL_by_powder}
\bibinfo{author}{\bibfnamefont{C.}~\bibnamefont{Gouedard}},
  \bibinfo{author}{\bibfnamefont{D.}~\bibnamefont{Husson}},
  \bibinfo{author}{\bibfnamefont{C.}~\bibnamefont{Sauteret}},
  \bibinfo{author}{\bibfnamefont{F.}~\bibnamefont{Auzel}}, \bibnamefont{and}
  \bibinfo{author}{\bibfnamefont{A.}~\bibnamefont{Migus}}, \bibinfo{journal}{J.
  Opt. Soc. Am. B} \textbf{\bibinfo{volume}{10}}, \bibinfo{pages}{2358}
  (\bibinfo{year}{1993}).

\bibitem[{\citenamefont{Letokhov}(1967)}]{Letokhov_NRA}
\bibinfo{author}{\bibfnamefont{V.}~\bibnamefont{Letokhov}},
  \bibinfo{journal}{Zh. Eksp. Teor. Fiz.} \textbf{\bibinfo{volume}{53}},
  \bibinfo{pages}{1442} (\bibinfo{year}{1967}).

\bibitem[{\citenamefont{Lawandy et~al.}(1994)\citenamefont{Lawandy,
  Balachandran, Gomes, and Sauvain}}]{Lawandy_Nature}
\bibinfo{author}{\bibfnamefont{N.~M.} \bibnamefont{Lawandy}},
  \bibinfo{author}{\bibfnamefont{R.~M.} \bibnamefont{Balachandran}},
  \bibinfo{author}{\bibfnamefont{A.~S.~L.} \bibnamefont{Gomes}},
  \bibnamefont{and} \bibinfo{author}{\bibfnamefont{E.}~\bibnamefont{Sauvain}},
  \bibinfo{journal}{Nature} \textbf{\bibinfo{volume}{368}},
  \bibinfo{pages}{436} (\bibinfo{year}{1994}).

\bibitem[{\citenamefont{Wiersma and Lagendijk}(1996)}]{Wiermsa_Diff_Gain}
\bibinfo{author}{\bibfnamefont{D.~S.} \bibnamefont{Wiersma}} \bibnamefont{and}
  \bibinfo{author}{\bibfnamefont{A.}~\bibnamefont{Lagendijk}},
  \bibinfo{journal}{Phys. Rev. E} \textbf{\bibinfo{volume}{54}},
  \bibinfo{pages}{4256} (\bibinfo{year}{1996}).

\bibitem[{\citenamefont{Cao et~al.}(1999)\citenamefont{Cao, Zhao, Ho, Seelig,
  Wang, and Chang}}]{Cao_RL_Action_Semiconductor}
\bibinfo{author}{\bibfnamefont{H.}~\bibnamefont{Cao}},
  \bibinfo{author}{\bibfnamefont{Y.~G.} \bibnamefont{Zhao}},
  \bibinfo{author}{\bibfnamefont{S.~T.} \bibnamefont{Ho}},
  \bibinfo{author}{\bibfnamefont{E.~W.} \bibnamefont{Seelig}},
  \bibinfo{author}{\bibfnamefont{Q.~H.} \bibnamefont{Wang}}, \bibnamefont{and}
  \bibinfo{author}{\bibfnamefont{R.~P.~H.} \bibnamefont{Chang}},
  \bibinfo{journal}{Phys. Rev. Lett.} \textbf{\bibinfo{volume}{82}},
  \bibinfo{pages}{2278} (\bibinfo{year}{1999}).

\bibitem[{\citenamefont{{van der Molen}
  et~al.}(2007{\natexlab{a}})\citenamefont{{van der Molen}, Tjerkstra, Mosk,
  and Lagendijk}}]{Lagendijk_spatial_extent_RL}
\bibinfo{author}{\bibfnamefont{K.~L.} \bibnamefont{{van der Molen}}},
  \bibinfo{author}{\bibfnamefont{R.~W.} \bibnamefont{Tjerkstra}},
  \bibinfo{author}{\bibfnamefont{A.~P.} \bibnamefont{Mosk}}, \bibnamefont{and}
  \bibinfo{author}{\bibfnamefont{A.}~\bibnamefont{Lagendijk}},
  \bibinfo{journal}{Phys. Rev. Lett.} \textbf{\bibinfo{volume}{98}},
  \bibinfo{pages}{143901} (\bibinfo{year}{2007}{\natexlab{a}}).

\bibitem[{\citenamefont{{van der Molen}
  et~al.}(2007{\natexlab{b}})\citenamefont{{van der Molen}, Mosk, and
  Lagendijk}}]{VanderMolen_Several_RL}
\bibinfo{author}{\bibfnamefont{K.~L.} \bibnamefont{{van der Molen}}},
  \bibinfo{author}{\bibfnamefont{A.~P.} \bibnamefont{Mosk}}, \bibnamefont{and}
  \bibinfo{author}{\bibfnamefont{A.}~\bibnamefont{Lagendijk}},
  \bibinfo{journal}{Optics Communications} \textbf{\bibinfo{volume}{278}},
  \bibinfo{pages}{110} (\bibinfo{year}{2007}{\natexlab{b}}), ISSN
  \bibinfo{issn}{0030-4018}.

\bibitem[{\citenamefont{Conti and Fratalocchi}(2008)}]{Conti08PhC}
\bibinfo{author}{\bibfnamefont{C.}~\bibnamefont{Conti}} \bibnamefont{and}
  \bibinfo{author}{\bibfnamefont{A.}~\bibnamefont{Fratalocchi}},
  \bibinfo{journal}{Nat. Physics} \textbf{\bibinfo{volume}{4}},
  \bibinfo{pages}{794} (\bibinfo{year}{2008}).

\bibitem[{\citenamefont{Cao et~al.}(2000)\citenamefont{Cao, Xu, Zhang, Chang,
  Ho, Seelig, Liu, and Chang}}]{Cao_spatial_confinement}
\bibinfo{author}{\bibfnamefont{H.}~\bibnamefont{Cao}},
  \bibinfo{author}{\bibfnamefont{J.~Y.} \bibnamefont{Xu}},
  \bibinfo{author}{\bibfnamefont{D.~Z.} \bibnamefont{Zhang}},
  \bibinfo{author}{\bibfnamefont{S.-H.} \bibnamefont{Chang}},
  \bibinfo{author}{\bibfnamefont{S.~T.} \bibnamefont{Ho}},
  \bibinfo{author}{\bibfnamefont{E.~W.} \bibnamefont{Seelig}},
  \bibinfo{author}{\bibfnamefont{X.}~\bibnamefont{Liu}}, \bibnamefont{and}
  \bibinfo{author}{\bibfnamefont{R.~P.~H.} \bibnamefont{Chang}},
  \bibinfo{journal}{Phys. Rev. Lett.} \textbf{\bibinfo{volume}{84}},
  \bibinfo{pages}{5584} (\bibinfo{year}{2000}).

\bibitem[{\citenamefont{Fallert et~al.}(2009)\citenamefont{Fallert, Dietz,
  Sartor, Schneider, Klingshirn, and Kalt}}]{Fallert_coexistence_nature}
\bibinfo{author}{\bibfnamefont{J.}~\bibnamefont{Fallert}},
  \bibinfo{author}{\bibfnamefont{R.~J.~B.} \bibnamefont{Dietz}},
  \bibinfo{author}{\bibfnamefont{J.}~\bibnamefont{Sartor}},
  \bibinfo{author}{\bibfnamefont{D.}~\bibnamefont{Schneider}},
  \bibinfo{author}{\bibfnamefont{C.}~\bibnamefont{Klingshirn}},
  \bibnamefont{and} \bibinfo{author}{\bibfnamefont{H.}~\bibnamefont{Kalt}},
  \bibinfo{journal}{Nat Photon} \textbf{\bibinfo{volume}{3}},
  \bibinfo{pages}{279} (\bibinfo{year}{2009}), ISSN \bibinfo{issn}{1749-4885}.

\bibitem[{\citenamefont{Tureci et~al.}(2008)\citenamefont{Tureci, Ge, Rotter,
  and Stone}}]{Tureci:08}
\bibinfo{author}{\bibfnamefont{H.~E.} \bibnamefont{Tureci}},
  \bibinfo{author}{\bibfnamefont{L.}~\bibnamefont{Ge}},
  \bibinfo{author}{\bibfnamefont{S.}~\bibnamefont{Rotter}}, \bibnamefont{and}
  \bibinfo{author}{\bibfnamefont{A.~D.} \bibnamefont{Stone}},
  \bibinfo{journal}{Science} \textbf{\bibinfo{volume}{320}},
  \bibinfo{pages}{643} (\bibinfo{year}{2008}).

\bibitem[{\citenamefont{Haus}(2000)}]{Haus_Mode_Loking}
\bibinfo{author}{\bibfnamefont{H.}~\bibnamefont{Haus}},
  \bibinfo{journal}{Selected Topics in Quantum Electronics, IEEE Journal of}
  \textbf{\bibinfo{volume}{6}}, \bibinfo{pages}{1173} (\bibinfo{year}{2000}),
  ISSN \bibinfo{issn}{1077-260X}.

\bibitem[{\citenamefont{Kutz}(2006)}]{Kutz_MLSL}
\bibinfo{author}{\bibfnamefont{J.~N.} \bibnamefont{Kutz}},
  \bibinfo{journal}{SIAM Rev.} \textbf{\bibinfo{volume}{48}},
  \bibinfo{pages}{629} (\bibinfo{year}{2006}), ISSN \bibinfo{issn}{0036-1445}.

\bibitem[{\citenamefont{Gordon and Fischer}(2002)}]{Gordon_Fischer_PRL_2002}
\bibinfo{author}{\bibfnamefont{A.}~\bibnamefont{Gordon}} \bibnamefont{and}
  \bibinfo{author}{\bibfnamefont{B.}~\bibnamefont{Fischer}},
  \bibinfo{journal}{Phys. Rev. Lett.} \textbf{\bibinfo{volume}{89}},
  \bibinfo{pages}{103901} (\bibinfo{year}{2002}).

\bibitem[{\citenamefont{Picozzi and Haelterman}(2004)}]{Picozzi_Condesation}
\bibinfo{author}{\bibfnamefont{A.}~\bibnamefont{Picozzi}} \bibnamefont{and}
  \bibinfo{author}{\bibfnamefont{M.}~\bibnamefont{Haelterman}},
  \bibinfo{journal}{Phys. Rev. Lett.} \textbf{\bibinfo{volume}{92}},
  \bibinfo{pages}{103901} (\bibinfo{year}{2004}).

\bibitem[{\citenamefont{El-Dardiry et~al.}(2010)\citenamefont{El-Dardiry, Mosk,
  Muskens, and Lagendijk}}]{PhysRevA.81.043830}
\bibinfo{author}{\bibfnamefont{R.~G.~S.} \bibnamefont{El-Dardiry}},
  \bibinfo{author}{\bibfnamefont{A.~P.} \bibnamefont{Mosk}},
  \bibinfo{author}{\bibfnamefont{O.~L.} \bibnamefont{Muskens}},
  \bibnamefont{and}
  \bibinfo{author}{\bibfnamefont{A.}~\bibnamefont{Lagendijk}},
  \bibinfo{journal}{Phys. Rev. A} \textbf{\bibinfo{volume}{81}},
  \bibinfo{pages}{043830} (\bibinfo{year}{2010}).

\bibitem[{\citenamefont{Chabanov et~al.}(2003)\citenamefont{Chabanov, Zhang,
  and Genack}}]{PhysRevLett.90.203903}
\bibinfo{author}{\bibfnamefont{A.~A.} \bibnamefont{Chabanov}},
  \bibinfo{author}{\bibfnamefont{Z.~Q.} \bibnamefont{Zhang}}, \bibnamefont{and}
  \bibinfo{author}{\bibfnamefont{A.~Z.} \bibnamefont{Genack}},
  \bibinfo{journal}{Phys. Rev. Lett.} \textbf{\bibinfo{volume}{90}},
  \bibinfo{pages}{203903} (\bibinfo{year}{2003}).

\bibitem[{\citenamefont{Cao et~al.}(2003)\citenamefont{Cao, Jiang, Ling, Xu,
  and Soukoulis}}]{PhysRevB.67.161101}
\bibinfo{author}{\bibfnamefont{H.}~\bibnamefont{Cao}},
  \bibinfo{author}{\bibfnamefont{X.}~\bibnamefont{Jiang}},
  \bibinfo{author}{\bibfnamefont{Y.}~\bibnamefont{Ling}},
  \bibinfo{author}{\bibfnamefont{J.~Y.} \bibnamefont{Xu}}, \bibnamefont{and}
  \bibinfo{author}{\bibfnamefont{C.~M.} \bibnamefont{Soukoulis}},
  \bibinfo{journal}{Phys. Rev. B} \textbf{\bibinfo{volume}{67}},
  \bibinfo{pages}{161101} (\bibinfo{year}{2003}).

\bibitem[{\citenamefont{{van der Molen}
  et~al.}(2007{\natexlab{c}})\citenamefont{{van der Molen}, Tjerkstra, Mosk,
  and Lagendijk}}]{PhysRevLett.98.143901}
\bibinfo{author}{\bibfnamefont{K.~L.} \bibnamefont{{van der Molen}}},
  \bibinfo{author}{\bibfnamefont{R.~W.} \bibnamefont{Tjerkstra}},
  \bibinfo{author}{\bibfnamefont{A.~P.} \bibnamefont{Mosk}}, \bibnamefont{and}
  \bibinfo{author}{\bibfnamefont{A.}~\bibnamefont{Lagendijk}},
  \bibinfo{journal}{Phys. Rev. Lett.} \textbf{\bibinfo{volume}{98}},
  \bibinfo{pages}{143901} (\bibinfo{year}{2007}{\natexlab{c}}).

\end{thebibliography}

\end{document}